\documentclass[graybox]{svmult}
\usepackage{titlesec}
\usepackage[utf8]{inputenc}
\usepackage{float}
\usepackage{amsmath}
\usepackage{subfig}
\usepackage{graphicx}

\begin{document}
\titlerunning{Magnetically arrested flows to explain ULXs}
\authorrunning{Raha, Mukhopadhyay, Chatterjee}
\title*{Magnetically arrested advective accretion flows and jets/outflows around stellar mass black holes: Explaining hard state ULXs with GRMHD simulations}

\author{Rohan Raha, Banibrata Mukhopadhyay and Koushik Chatterjee}
\institute{Rohan Raha \at  Department of Physics, Indian Institute of Science, Bengaluru, \email{raharohan@iisc.ac.in}
\and Banibrata Mukhopadhyay \at Department of Physics, Indian Institute of Science, Bengaluru \email{bm@iisc.ac.in}
\and Koushik Chatterjee \at University of Maryland, College Park, the USA \email{kchatt@umd.edu}}

\maketitle

\textit{To be published in Astrophysics and Space Science Proceedings, titled "The Relativistic Universe: From Classical to Quantum, Proceedings of the International Symposium on Recent Developments in Relativistic Astrophysics", Gangtok, December 11-13, 2023: to felicitate Prof. Banibrata Mukhopadhyay on his 50th Birth Anniversary", Editors: S Ghosh \& A R Rao, Springer Nature}

\vspace{2cm}

\abstract{
An optically thin advective accretion disk is crucial for explaining the hard state of black hole sources. Using general relativistic magnetohydrodynamic (GRMHD) simulations, we investigate how a large-scale, strong magnetic field influences accretion and outflows/jets, depending on the field geometry, magnetic field strength, and the spin parameter of the black hole. We simulate a sub-Eddington, advective disk-outflow system in the presence of a strong magnetic field, which likely remains in the hard state. The model simulations based on HARMPI successfully explain ultra-luminous X-ray sources (ULXs) in the hard state, typically observed with luminosities ranging from  $10^{39}$ - $10^{40}$ ergs s$^{-1}$. Our simulations generally describe the bright, hard state of stellar-mass black hole sources without requiring a super-Eddington accretion rate. This work explores the characteristics of ULXs without invoking intermediate-mass black holes. The observed high luminosity is attributed to the energy stored in the strong magnetic fields, which can generate super-Eddington luminosity. The combined energy of the matter and magnetic field leads to such significant luminosity.}

\section{Introduction}
Accretion disks surrounding black holes predominantly involve rotating gas, converting gravitational energy into radiation with high efficiency. This process is recognized as the key power source for various luminous astronomical systems, from quasars and active galactic nuclei (AGNs) with supermassive black holes to X-ray binaries (XRBs) with stellar-mass black holes. It is widely acknowledged that magnetic fields are present in these accretion disks and play a significant role in altering their physical states. Therefore, it is expected that magnetic fields contribute to numerous dynamic processes in these disks. For instance, magnetic stress could replace viscous stress as explained by the standard model \cite{1} proposed by Shakura and Sunyaev.

Accretion flows around black holes are generally divided into two categories: cold and hot flows. Cold flows consist of cool, optically thick gas, which typically occurs at higher mass accretion rates. This includes the standard thin disk producing multi-color blackbody radiation at sub-Eddington rates, as well as the slim disk at super-Eddington rates. Such flows are associated with bright objects like AGNs near the Eddington luminosity and soft-state black hole XRBs. On the other hand, hot flows can reach virial temperatures, are optically thin, and occur at lower accretion rates, featuring advection \cite{2}, \cite{3}. These flows generally have lower radiative efficiency compared to thin disks, and their efficiency decreases as the accretion rate drops. 

Black hole accretion disks transition through various X-ray spectral states, is largely based on their X-ray luminosity and spectral characteristics \cite{4}. 
The most common states are the high/soft (HS) and low/hard (LH) states \cite{4,5}, which have been the focus of extensive research. These states are believed to reflect different accretion geometries depending on the mass accretion rates and the density of matter.

However, ULXs in the hard state pose a significant scientific puzzle. These sources exhibit luminosities surpassing the Eddington limit for stellar-mass black holes, yet they persist in the hard state—a condition typically linked to lower accretion rates. This apparent contradiction challenges our current understanding. The Eddington luminosity, a crucial concept in this context, is mathematically expressed as:
\begin{equation}
L_{Edd}=\frac{4\pi cGM_{BH}}{\kappa_{es}}\approx1.4\times 10^{38}\left(\frac{M_{BH}}{M_{\odot}}\right) \text{erg}\,\,\text{s}^{-1},
\end{equation}
In this equation, $M_{BH}$ represents the black hole's mass, $G$ is Newton's gravitational constant, $c$ denotes the speed of light, and $\kappa_{es}$ stands for the electron scattering opacity.

A ULX in the hard state must be associated with an optically thin, advective accretion disk \cite{6}. To explore ULXs in this state, we analyze how a strong, large-scale magnetic field in the accretion disk, outflows or jets, may transport angular momentum, primarily through magnetic shear rather than $\alpha$-viscosity, depending on the magnetic field configuration and plasma-$\beta$ parameter. Interestingly, while the disk becomes thermally unstable at higher accretion rates, it regains stability when stronger magnetic fields are present \cite{7}. 
It was previously theorized \cite{6} \cite{8} \cite{9} \cite{10} \cite{11} that magnetically arrested advective accretion flow (MA-AAF) in an optically thin environment holds critical significance in explaining ULXs and other phenomena.
This research confirms these behaviors through numerical simulations. 

In this paper and a companion paper \cite{12}, we simulate a 
sub-Eddington, disk-outflow symbiotic model in the advective regime (hence hard state; though no radiation physics included) in the presence of large-scale strong magnetic fields. Numerical simulations of a sub-Eddington, disk-outflow model in the advective regime with strong magnetic fields aim to explain hard-state ULXs typically observed with luminosities of $10^{39}-10^{40}~\text{ergs}\,\text{s}^{-1}$ (e.g., NGC 1365, \cite{14}; Antennae X-11, X-16, X-42,
X-44, \cite{13};  M82 X42.3+59, \cite{16};
M99 X1, \cite{15}). Our model can also account for bright, hard states of stellar-mass black hole sources without requiring super-Eddington accretion rates or intermediate-mass black holes.

\section{General Relativistic Magnetohydrodynamic (GRMHD) framework}
The behavior of accretion flows is largely governed by the interplay of non-Maxwellian plasma with intense gravitational and electromagnetic fields, and potentially significant radiation. We model this plasma as a fluid, necessitating the use of GRMHD equations. Our study employs HARMPI \cite{17}, a high-performance parallel GRMHD code, to analyze these flows. HARMPI utilizes a robust numerical approach, combining conservative methods, shock-capture capabilities, and stability, founded on the MUSCL scheme \cite{18} and incorporating HLL fluxes \cite{19}. We concentrate on resolving the GRMHD momentum conservation equations and the continuity equation within specific constraints, applying appropriate boundary and initial conditions for an accretion disk. This approach enables temporal evolution of the solution. Using HARMPI, we investigate GRMHD solutions for radiatively inefficient, magnetized accretion flows surrounding a rotating black hole, characterized by the Kerr metric in Kerr-Schild coordinates.

The first governing equation that describes the accretion flow is the conservation of particle number
\begin{equation}
    (nu^\mu)_{;\mu}=0
\end{equation}
where $n$ is the particle number density and $u^\mu$ is the 4-velocity.
By replacing $n$ with mass density $\rho=mn$, when $m$ being the mass of a particle, we can rewrite the above equation in a coordinate basis
\begin{equation}
    \frac{1}{\sqrt{-g}}\partial_\mu(\sqrt{-g}\rho u^\mu)=0
\end{equation}
which reduces to
\begin{equation}
    \partial_t(\sqrt{-g}\rho u^t)=-\partial_i(\sqrt{-g}\rho u^i)
\end{equation}
where $i$ denotes spatial index.
The next set of equations is the conservation of energy and momentum, given by
\begin{equation}
    (T^\mu_\nu)_{;\mu}=0
\end{equation}
where $T^\mu_\nu$ is the stress-energy tensor. The equation in the coordinate basis reduces to
\begin{equation}
    \partial_t(\sqrt{-g} T^t_\nu)=-\partial_i(\sqrt{-g} T^i_\nu)+\sqrt{-g}T^\kappa_\lambda \Gamma^\lambda_{\nu\kappa}
\end{equation}
where $\Gamma^\lambda_{\nu\kappa}$ is the affine connection.

The hydrodynamic part of stress-energy tensor for a perfect fluid with an electromagnetic field is given by 
\begin{equation}
    T^{\mu\nu}_{HD}=(\rho+u+p)u^\mu u^\nu +p g^{\mu\nu}
\end{equation}
(where $u$ is the internal energy density of the gas and $p$ is the gas pressure; for ideal MHD they are related by $p=(\gamma-1)u$ with $\gamma$ being polytropic constant)
and the electromagnetic part is given by
\begin{equation}
    T^{\mu\nu}_{EM}=F^{\mu\alpha}F^\nu_\alpha-\frac{1}{4}g^{\mu\nu}F^{\alpha\beta}F_{\alpha\beta}
\end{equation}
where $F^{\alpha\beta}$ is the electromagnetic tensor.
Using the ideal MHD approximation, $\mathbf{E}+\mathbf{v}\mathbf{\times}\mathbf{B}=0$ as the fluid has very high conductivity. Hence, the Lorentz force on the plasma vanishes, which is equivalent to
\begin{equation}
\label{eq:idealcond}
    u_\mu F^{\mu\nu}=0.
\end{equation}
The magnetic field in the fluid frame is defined as 
\begin{equation}
    b^\mu=\frac{1}{2}\epsilon^{\mu\nu\kappa\lambda}u_\nu F_{\kappa\lambda}
\end{equation}
where $\epsilon$ is the Levi-Civita symbol. Using $b_\mu u^\mu=0$ we get the electromagnetic part of the stress energy tensor as 
\begin{equation}
    T^{\mu\nu}_{EM}=b^2 u^\mu u^\nu +\frac{1}{2} b^2 g^{\mu\nu}-b^\mu b^\nu.
\end{equation}
Therefore, the total stress-energy tensor is given by
\begin{equation}
    T^{\mu\nu}=(\rho+u+p+b^2)u^\mu u^\nu +(p+\frac{1}{2}b^2) g^{\mu\nu}-b^\mu b^\nu.
\end{equation}
The electromagnetic field evolution is given by the Maxwell equation 
\begin{eqnarray}
\label{eqn:ind}
    &F_{\mu\nu,\lambda}+F_{\lambda\mu,\nu}+F_{\nu\lambda,\mu}=0 \\ 
    &F^{*\mu\nu}_{;\nu}=0
\end{eqnarray}
where $F^{*\mu\nu}$ is the dual of the electromagnetic field tensor given by
\begin{equation}
    F^{*\mu\nu}=\frac{1}{2}\epsilon^{\mu\nu\kappa\lambda}F_{\kappa\lambda}.
\end{equation}

\section{Simulation Setup}
We use modified Kerr–Schild coordinates $(x_0, x_1, x_2, x_3)$ given by
\begin{equation}
    t=x_0,\, \phi=x_3, \, r=e^{x_1}, \, \theta=\pi x_2+\frac{1}{2}(1-h)\sin(2\pi h x_2).
\end{equation}
Our 2D simulation
grid extends from $r = r_{in} = 0.6 + 0.4r_H$ to $10^3r_g$ with $r_H$ being the radius of event horizon and $r_g$ is the gravitational radius given by $r_g = GM_{BH}/c^2$. The grid resolution used is $N_r \times N_\theta \times N_\phi \equiv 256 \times 256 \times 1$. We apply outflowing
radial boundary conditions (BCs), transmissive polar BCs,
and periodic BCs in the azimuthal direction.

Our simulation starts with a black hole surrounded by a standard Fishbone-Moncrief (FM) torus  \cite{20}. 
 We use an
ideal gas equation of state with the gas pressure $p =(\gamma-1)u$, where $\gamma = 5/3$. We evolve the initial magnetized FM torus from time $t=0$ to $t=20000 r_g/c$.  

We initialize our model with a poloidal magnetic field by
applying a toroidal magnetic vector potential given by, $A_\phi= max(q, 0)$ \cite{21}, where 
\begin{equation}
    q=\frac{\rho}{\rho_{max}}-0.2
\end{equation}
and $\rho$ is the rest-mass gas density. The
magnetic field strength in the initial setup is normalized by
setting $max(p_{gas})/max(p_{mag}) = 0.1$, where $p_{mag} = b^2/2$, is the
magnetic pressure. Here $max$ denotes the maximum of the respective quantity.

All quantities are computed from the results produced in every $\sim10r_g/c$.
For quantities $Q$, averages over space ($<Q>$) and time ($[Q]_t$) are performed directly on $Q$.  

The mass accretion rate as a function of radius ($r$) is calculated as 
\begin{equation}
    \dot{M}(r)=-\int \int \sqrt{-g} \rho u^r d\theta d\phi
\end{equation}
and the radial energy flux is given by
\begin{equation}
\label{eq:energy}
    \dot{E}(r)=\int \int \sqrt{-g} T^r_t d\theta d\phi.
\end{equation}
The net outflow efficiency is then given by 
\begin{equation}
\label{eqn:eff}
    \eta=\frac{\dot{M}-\dot{E}}{\dot{M}}=\frac{P_{out}}{\dot{M}}.
\end{equation}
This refers to the power of outflows that escape to infinity normalized by the mass accretion rate.
Positive efficiency values correspond to energy extraction from the system through jets or outflows.
We report the outflow power/efficiency time averaged from $t=15,000 r_g/c$ to $t=20,000 r_g/c$.

\section{Results}

We aim to investigate, through computational models, how the magnetic field and black hole spin influence the accretion process. We investigate the mechanism through which material is pulled into the black hole via magnetic fields associated with the accretion disk. To validate our findings, we have conducted simulations based on previously established parameters \cite{21}. We have began with a magnetized torus, using a black hole with a spin parameter of $a=0.9375$ and an initial plasma-$\beta$=100. We also compare our results with the GRMHD simulations carried out using the publicly available code BHAC (Black Hole Accretion Code) \cite{22}, for the same parameters,  published in the present volume \cite{12}. We evolve the disk in time till $t=20,000 r_g/c$. We compute the power based on equation (\ref{eqn:eff}).

Throughout, when we report the results in CGS units, we consider a stellar mass black hole of mass $M_{BH}=20 M_\odot$ with total mass accretion rate, $\dot{M} = 0.05 \dot{M}_{Edd}$, where $\dot{M}_{Edd} = L_{Edd}/\eta c^2 = 1.39 \times 10^{18}(M_{BH}/M)$ g s$^{-1}$, considering radiative efficiency $\eta = 0.1$. 

\begin{figure}[H]%
    \centering
    \subfloat[\centering]{{\includegraphics[height=7cm,width=10cm]{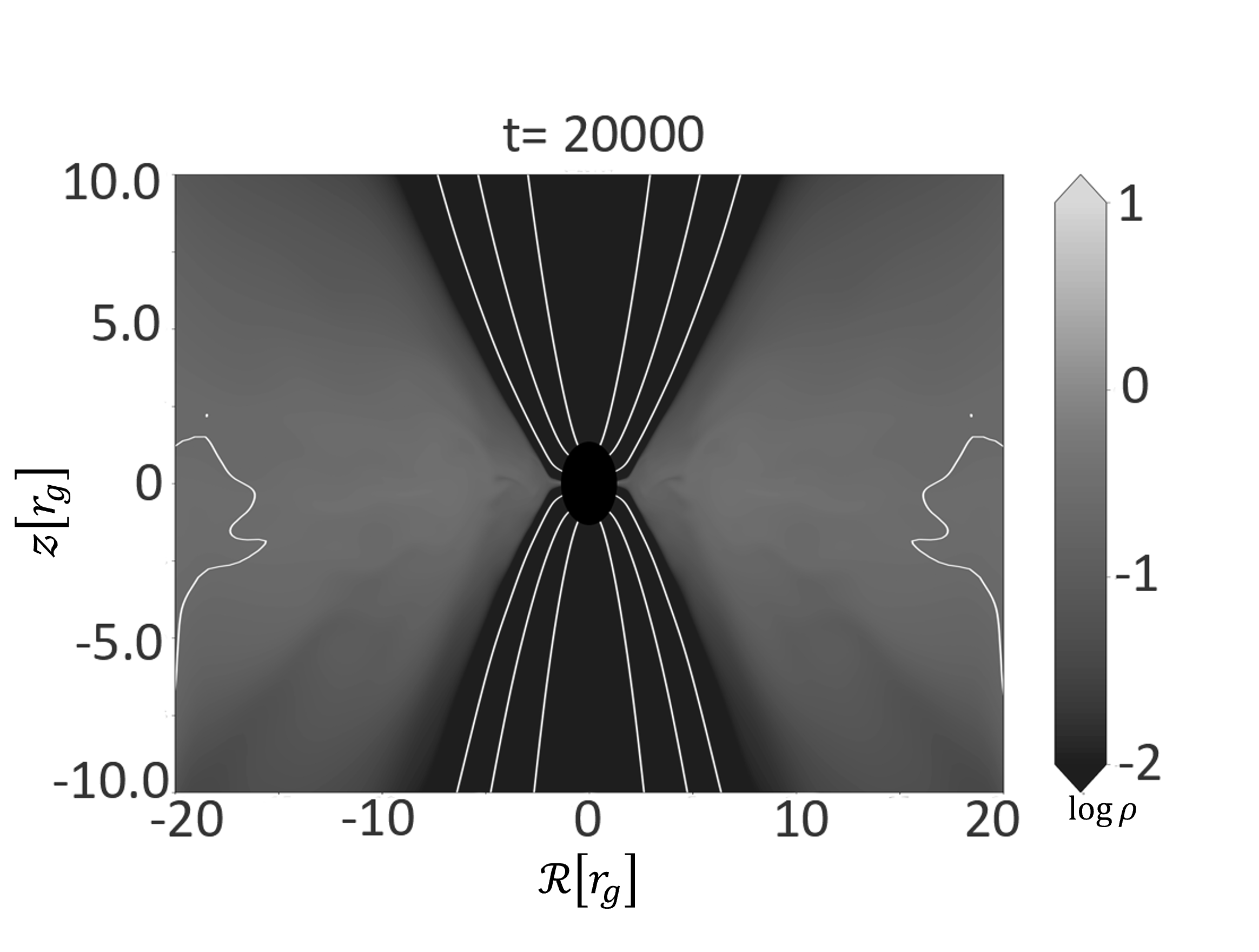} }}%
    \qquad
    \subfloat[\centering]{{\includegraphics[height=5.5cm,width=8.5cm]{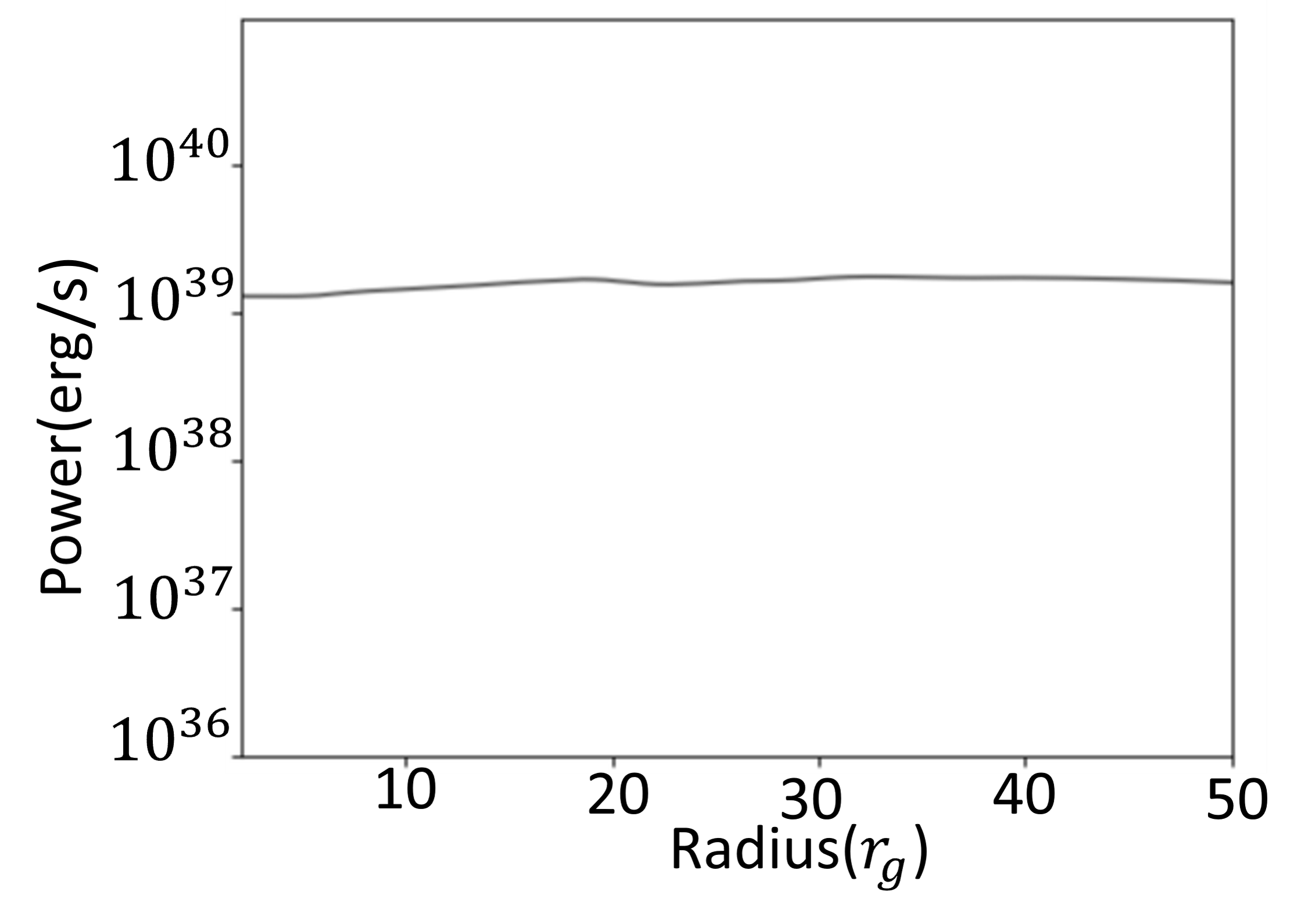} }}%
    \caption{(a) The density contour at the flow vertical plane in log-scale at time $t=20,000 r_g/c$ with the white lines indicating magnetic field lines, for $a=0.9375$, initial plasma-$\beta=100$, (b) the corresponding outflow power in CGS units. }
    \label{fig:common}%
\end{figure}
.

 Figure \ref{fig:common} shows the snapshot of density in log-scale at time $t=20,000 r_g/c$ along with the outflow power. This shows an efficiency of around $1.2$, as was demonstrated earlier \cite{21}. A figure with similar parameters is presented in other work exploring BHAC \cite{21} for comparison.

\begin{figure}[H]
    \centering
    \includegraphics[height=7cm,width=10cm]{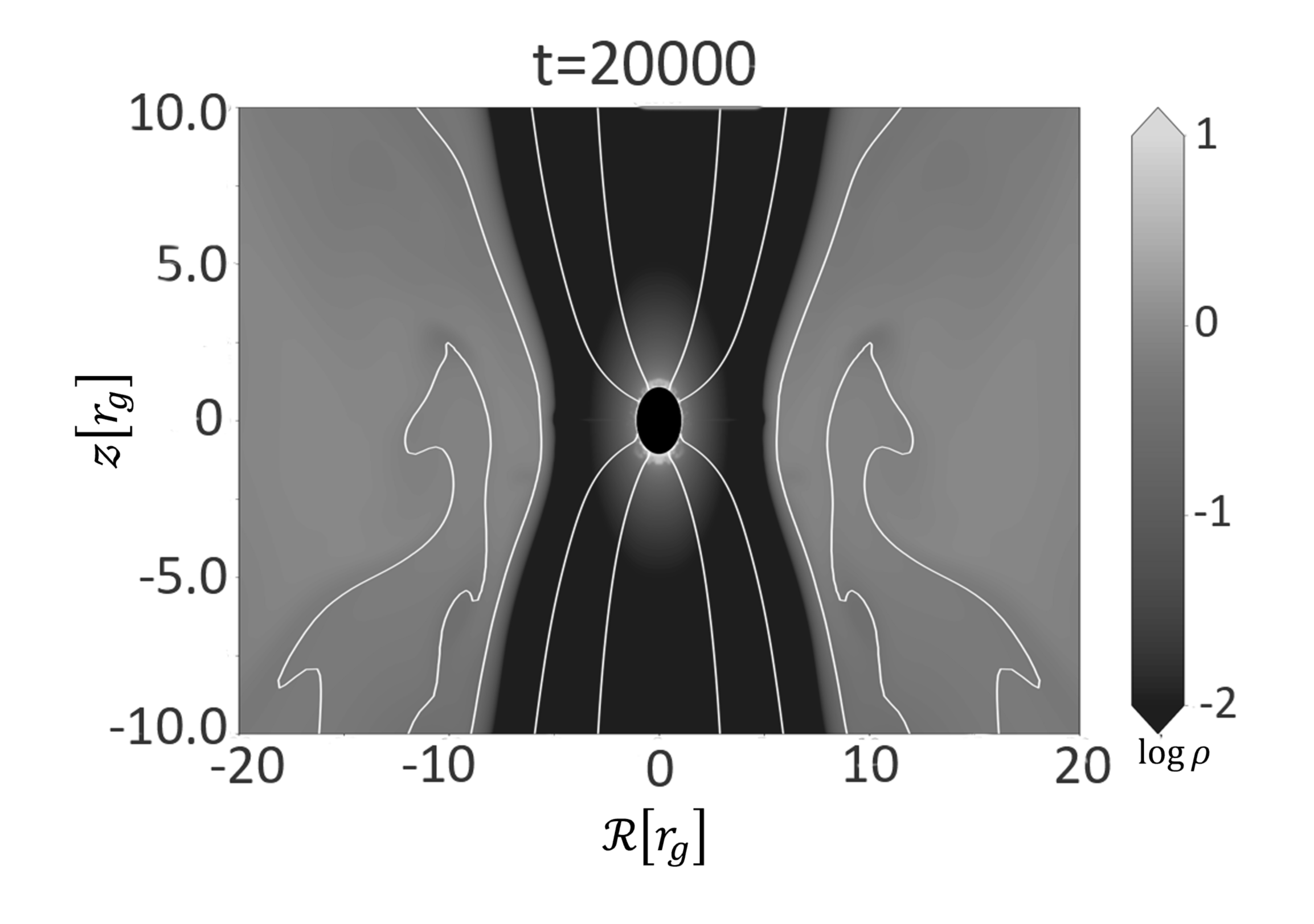}%
        \caption{Density contour in log-scale at time $t=20,000 r_g/c$ with the white lines indicating magnetic field lines for the case $a=0.998$, initial plasma-$\beta=0.1$.}%
    \label{fig:density1}
\end{figure}

\begin{figure}[H]
    \centering
    \includegraphics[height=7cm,width=10cm]{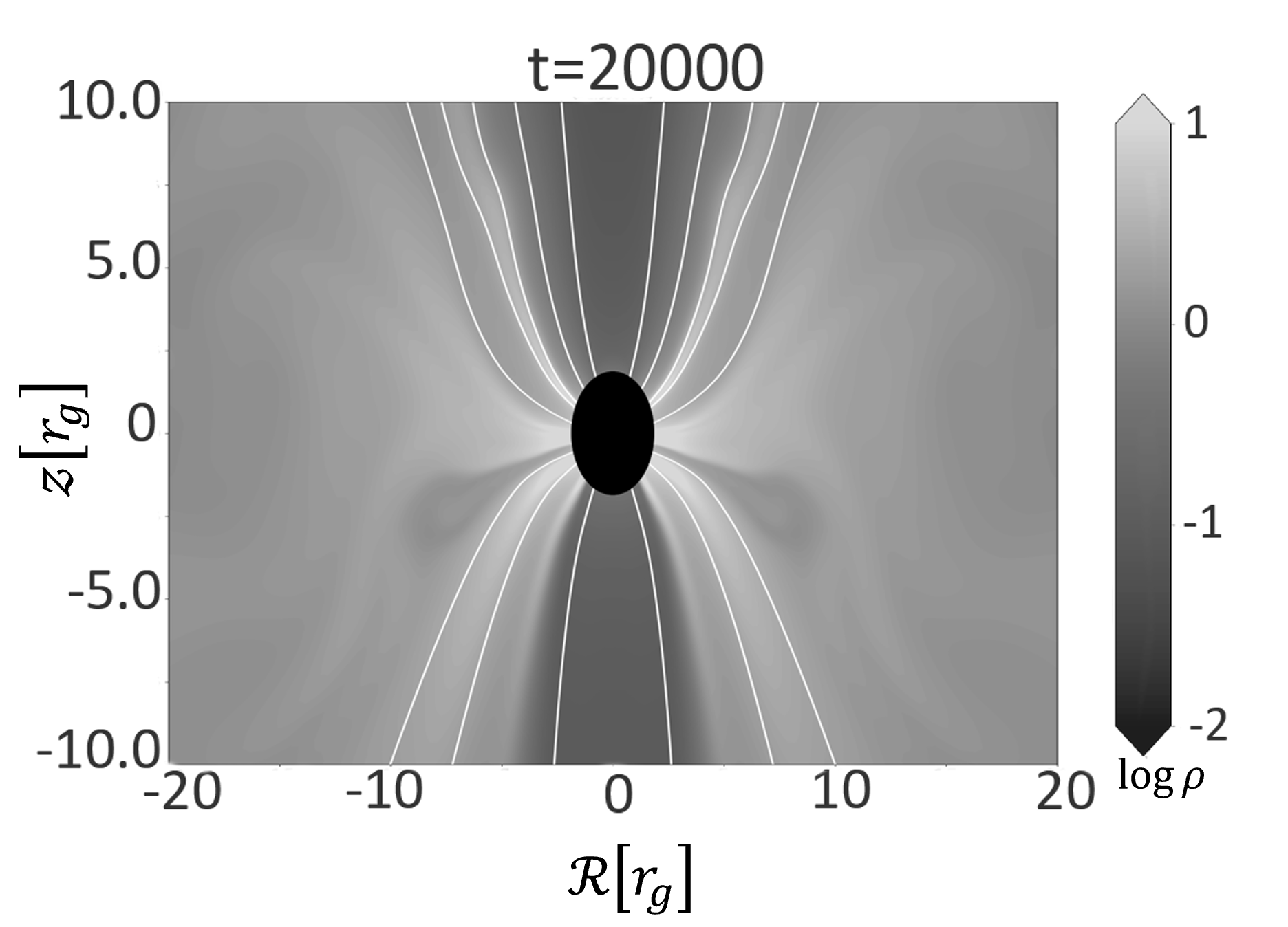}%
        \caption{Density contour at the flow vertical plane in log-scale at time $t=20,000 r_g/c$ with the white lines indicating magnetic field lines for $a=0.5$, initial plasma-$\beta=0.1$.}%
    \label{fig:density2}
\end{figure}

\begin{figure}[H]
    \centering
    \includegraphics[height=7cm,width=11cm]{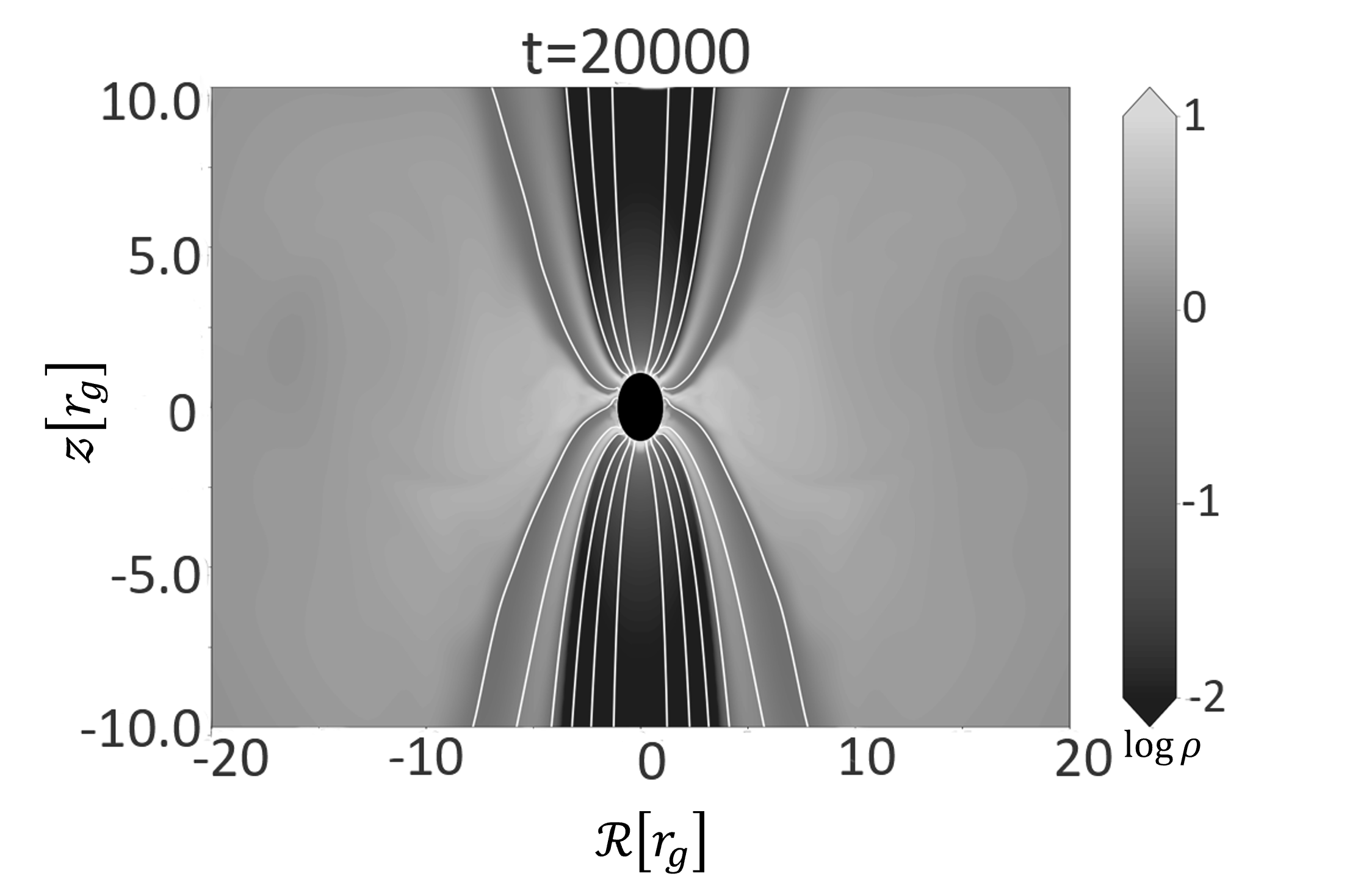}%
        \caption{Density contour at the flow vertical plane in log-scale at time $t=20,000 r_g/c$ with the white lines indicating magnetic field lines for $a=0.998$, initial plasma-$\beta=1$.}%
    \label{fig:density4}
\end{figure}

We further run simulations for two different spin parameters $a=0.5$ and $0.998$, i.e., low, and high spins, respectively, and start with an initial plasma-$\beta$=0.1. To capture the effect of the magnetic field, we also run simulations with $a=0.998$ and initial plasma-$\beta$=1, which corresponds to a lower magnetic field case. Figures \ref{fig:density1}, \ref{fig:density2} and \ref{fig:density4} show snapshots of the simulation results
of accretion disk for the three cases, respectively, where they show the inflow-outflow structure
at an evolved time $t=20,000 r_g/c$.

During the accretion process, the matter pulls the magnetic field lines inward because the magnetic flux is nearly frozen into the matter due to the ideal MHD condition. As the gas draws the magnetic field inward and interacts with the black hole's spin, the surrounding spacetime also gets dragged along. Moreover, the disk's rotation twists the magnetic field lines around the black hole \cite{23}. Consequently, the magnetic field lines are carried along with the space around the black hole. This causes the field lines to form a helical shape around the black hole, moving outward in a wave-like fashion. This wave transports energy and angular momentum away from the black hole through outflows and jets \cite{24}. Therefore, a black hole with a higher spin can more effectively drag both the matter and magnetic field, leading to faster inward motion of the material due to increased transport by magnetic shear. This mechanism generates more potent jets compared to scenarios involving lower spin rates. As compared in Figures \ref{fig:density1}, \ref{fig:density2} and \ref{fig:density4}, for the $a=0.998$ and initial plasma-$\beta$=0.1 case, the magnetic field forms a barrier near the black hole, thus forming a magnetically arrested disk through which matter diffuses after some time. As the matter brings in more magnetic flux due to flux freezing, it forms a barrier again, and this process goes on. However, in the case of $a=0.5$ and initial plasma-$\beta$=0.1, in Figure \ref{fig:density2}, the spin is not high enough to drag the magnetic field as efficiently to form a barrier. In Figure \ref{fig:density4}, for the case of $a=0.998$ and initial plasma-$\beta$=1, although the spin is high, the initial magnetic field supply is not high enough to produce magnetic barriers. They produce outflows that are weaker than the case with initial plasma-$\beta$=0.1.

The possible explanations for these jets and outflows have been proposed by  Blandford–Znajek (BZ) \cite{25} and Blandford–Payne (BP) \cite{26}
mechanisms. It has been explained, how the toroidal magnetic field helps launch powerful outflows, which are then collimated by the poloidal magnetic field. 

The Blandford-Znajek model suggests that the power of jets is determined by the magnetic flux and the black hole's spin, expressed as $P\propto \phi_{BH}^2 a^2$, where $\phi_{BH}$ is the poloidal magnetic flux accumulated near the black hole and $a$ is the spin of the black hole. Consequently, a higher black hole spin leads to matter falling inward at a greater velocity due to increased transport by magnetic shear, resulting in stronger outflows compared to a lower spin scenario.

\begin{figure}[H]
    \centering
    \includegraphics[width=8cm,height=6cm]{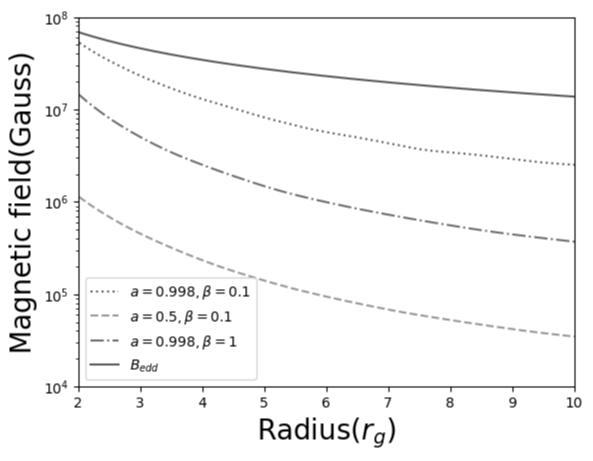}%
    \caption{Magnetic field in Gauss for (i) $a=0.998$, initial plasma-$\beta=0.1$, (ii) $a=0.5$, initial plasma-$\beta=0.1$, and (iii) $a=0.998$, initial plasma-$\beta=1$.}%
    \label{fig:mag}
\end{figure}

Figure \ref{fig:mag} shows the time averaged magnetic field for the three cases. It demonstrates how the magnetic field evolves for different spins. As discussed earlier, the higher spinning black hole can gather a larger magnetic field due to an increased frame dragging effect and hence has more magnetic flux at its disposal to launch more powerful outflows.  High polarization in the hard state of Cyg X-1 with $M\sim 14 M_\odot$ and related jet suggest magnetic field $B\sim 10^6 G$ at the base of jet (i.e. disk) \cite{27}, which is recovered from the simulations. 
Near black holes, the magnetic field's influence becomes predominant due to the process of flux freezing, which results from the inward movement of magnetic flux during accretion. However, there's a ceiling on the magnetic flux that can accumulate around a black hole. This upper bound on magnetic field intensity is established by equating the magnetic field's energy density to that of the accreting matter, which generates the Eddington luminosity in the vicinity of the black hole. This limit is referred to as the Eddington magnetic field \cite{8} and can be described by the expression $B_{Edd}\sim10^4G(M/10^9M_\odot)^{-1/2}$. Due to the super-Eddington magnetic field very close to the black hole in the two cases, i.e.,  $a=0.998$, initial plasma-$\beta=0.1$, and $a=0.998$, initial plasma-$\beta=1$, these systems can produce very high energy outflows as the high magnetic field can throw out matter to produce energetic outflows.

We calculate the outflow power from the efficiency based on the definitions
\begin{equation}
\eta=\frac{[\dot{M}]_t-[\dot{E}]_t}{[\dot{M}]_{t}}.\\
\end{equation}
This is equivalent to the definition in equation (\ref{eqn:eff}) but here we have normalized the power by time averaged mass accretion rate $[M]_{t}$ and time is averaged from $t=15,000 r_g/c$ to $t=20,000 r_g/c$.

\begin{figure}[H]
    \centering
    \centering{{\includegraphics[scale=0.5]{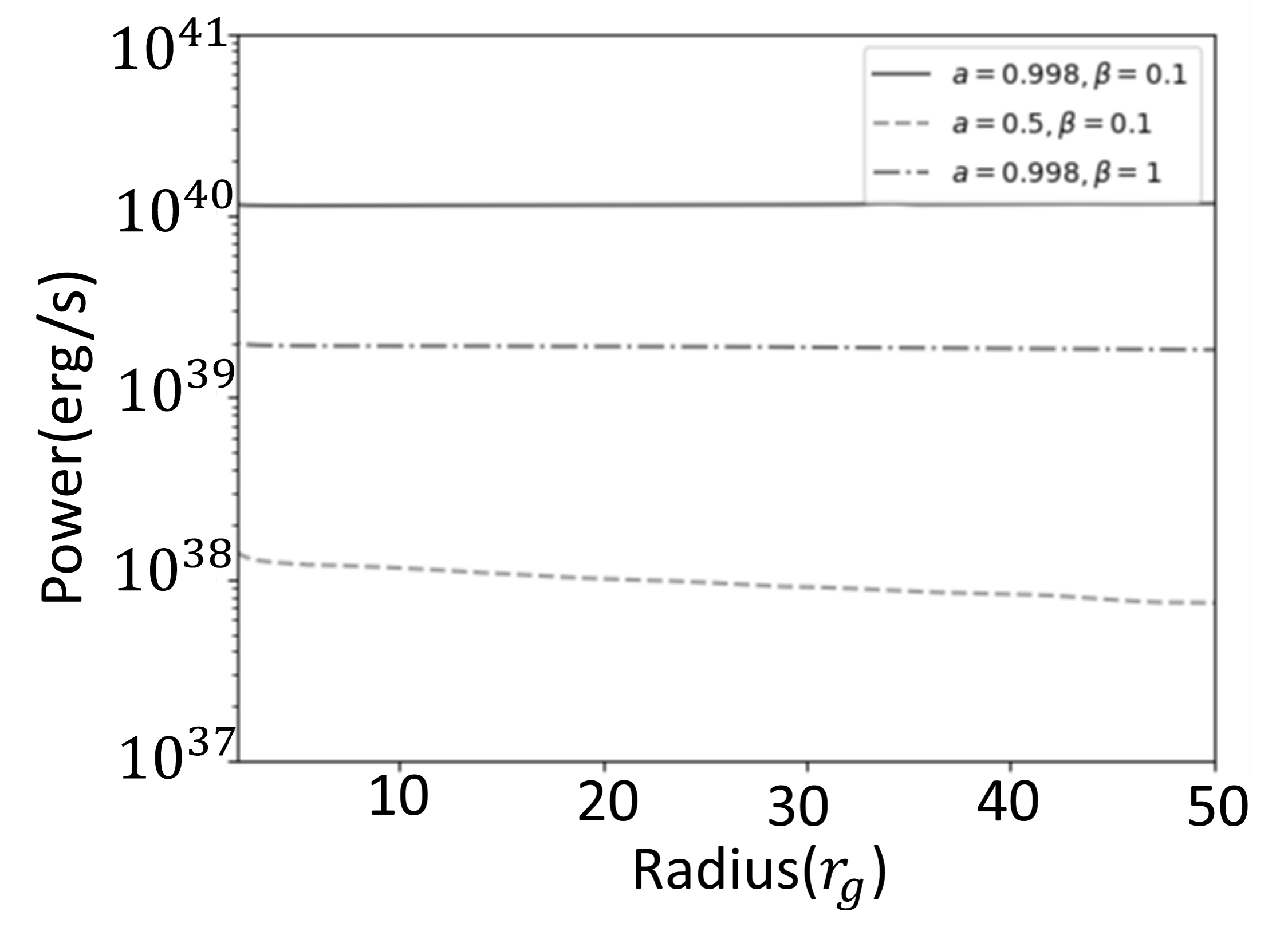} }}%
    \caption{Outflow Power in erg/s: (a) Power for (i) $a=0.998$, initial plasma-$\beta=0.1$, (ii) $a=0.5$, initial plasma-$\beta=0.1$, and (iii) $a=0.998$, initial plasma-$\beta=1$}%
    \label{fig:Power}
\end{figure}

Figure \ref{fig:Power} shows the outflow power for the three cases discussed above. As expected, the outflow power increases with the spin of the black hole and also increases with the magnetic field strength (shown in Figure \ref{fig:mag}). We see a direct correlation between the magnetic field and outflow power, because the outflow power increases as we have a higher magnetic field very close to the black hole horizon. As predicted from Blandford-Znajek model discussed earlier, we observe a higher outflow power with increased magnetic field strength and higher black hole spin.

\section{Summary}

We have conducted simulations of a sub-Keplerian, magnetized, advective accretion disk, featuring both inflowing and outflowing matter. This disk, being sub-Keplerian and advective, is also optically thin, with inflow and outflow present. The presence of strong magnetic fields generates effective magnetic forces, which act similarly to viscous forces. These forces facilitate the transport of angular momentum both within the disk and through the outflow. To sustain hard X-rays, the accretion rate needs to have a strict upper limit, which is why the system remains non-radiative in nature with a lower accretion rate. However, radiative GRMHD simulations could provide a more accurate picture of the dynamics of accretion flows. We have computed the total outflow power from the system and have found that to generate highly energetic outflows, a high spin of the black hole, along with a substantial magnetic field, is required. The black hole can drag in the magnetic field, launching powerful jets.

The presence of strong magnetic fields leads to the formation of a highly energetic disk and strong outflows or jets. We observed magnetic field strengths of up to $10^6$ Gauss at the bases of jets, consistent with observations of systems like Cyg X-1. Our simulations show a direct correlation between the magnetic field strength and outflow power, as predicted by the Blandford-Znajek model. Higher black hole spin and stronger magnetic fields result in more powerful outflows. For cases with high spin ($a = 0.998$) and strong initial magnetic fields (plasma-$\beta = 0.1$), we observed the formation of magnetically arrested disks, where magnetic barriers periodically form and dissipate.

The outflow power achieved in our simulations reaches levels consistent with observed ULXs, with luminosities in the range of $10^{39} - 10^{40}$ ergs/s, without requiring super-Eddington accretion rates or intermediate-mass black holes. This model can be applied to explain ULXs in their hard state. These ULXs can be described as magnetically dominated, advective, sub-Keplerian accretion flows, or MA-AAF, surrounding a stellar-mass black hole with a high spin. Our simulations demonstrate that the combination of strong magnetic fields and high black hole spin can produce the high luminosities observed in ULXs while maintaining the characteristics of the hard state.

Future work could include incorporating radiative GRMHD simulations to provide a more complete picture of the accretion flow dynamics, exploring a wider range of initial conditions, including different black hole masses and accretion rates, investigating the long-term stability of these magnetically arrested flows and their implications for ULX variability, and comparing simulated spectra with observed ULX spectra to further validate the model. These findings contribute to our understanding of ULXs and suggest that some of these enigmatic sources may be explained by extreme magnetic fields around rapidly spinning stellar-mass black holes, rather than requiring intermediate-mass black holes or super-Eddington accretion.


\clearpage

\end{document}